\documentclass[12pt,hidelinks]{article}
\usepackage{ulem}
\usepackage[utf8]{inputenc}
\usepackage{amsmath, amsfonts}
\usepackage[margin=1.0in]{geometry}
\usepackage{algorithm}
\usepackage[noend]{algpseudocode}
\usepackage{graphics, graphicx}
\usepackage{authblk}
\usepackage{xcolor}
\usepackage{hyperref}
\usepackage{multirow}
\usepackage{caption}
\usepackage{bm}
\usepackage{natbib}
\usepackage{booktabs}
\usepackage{appendix}
\usepackage{tcolorbox}
\usepackage{longtable}
\usepackage{afterpage}
\usepackage{pdflscape}
\usepackage{capt-of}
\usepackage{lscape}
\usepackage{subcaption}
\usepackage{mathtools}
\usepackage{setspace}
\usepackage[nomarkers,heads,tablesfirst]{endfloat}
\usepackage{epstopdf}
\AtBeginDelayedFloats{}

\DeclareMathOperator{\EX}{\mathbb{E}}

\title{Bayesian Varying-Effects Vector Autoregressive Models for Inference of Brain Connectivity Networks and Covariate Effects in Pediatric Traumatic Brain Injury}

\author[1,*]{Yangfan Ren}
\author[2,*]{Nathan Osborne}
\author[3]{Christine B. Peterson}
\author[4]{Dana M. DeMaster}
\author[4]{Linda Ewing-Cobbs}
\author[1**]{Marina Vannucci}
\affil[1]{Department of Statistics, Rice University, Houston, TX}
\affil[2]{Intuit, San Diego, CA}
\affil[3]{Department of Biostatistics, The University of Texas MD Anderson Cancer Center, Houston, TX}
\affil[4]{Department of Pediatrics, Children's Learning Institute, University of Texas Health Science Center, Houston, TX}
\affil[*]{These authors contributed equally}
\affil[**]{Corresponding author: marina@rice.edu}
\begin{document}

\maketitle

\begin{abstract}
In this paper, we develop an analytical approach for estimating brain connectivity networks that accounts for subject heterogeneity. More specifically, we consider a novel extension of a multi-subject Bayesian vector autoregressive model that estimates group-specific directed brain connectivity networks and accounts for the effects of covariates on the network edges. We adopt a flexible approach, allowing for (possibly) non-linear effects of the covariates on edge strength via a novel Bayesian nonparametric prior that employs a weighted mixture of Gaussian processes. For posterior inference, we achieve computational scalability by implementing a variational Bayes scheme. Our approach enables simultaneous estimation of group-specific networks and selection of relevant covariate effects. We show improved performance over competing two-stage approaches on simulated data.  We apply our method on resting-state fMRI data from children with a history of traumatic brain injury and healthy controls to estimate the effects of age and sex on the group-level connectivities.  Our results highlight differences in the distribution of parent nodes. They also suggest alteration in the relation of age, with peak edge strength in children with \textcolor{black}{traumatic brain injury (TBI)}, and differences in effective connectivity strength between males and females. \\



\end{abstract}

\section{Introduction}
Functional magnetic resonance imaging (fMRI) measures blood-oxygenation-level-dependent (BOLD) contrast, which reflects the difference in magnetization between oxygenated and deoxygenated blood arising from patterns in cerebral blood flow. Changes in BOLD response are treated as a proxy for changes in neurological activity.  This technique remains one of the most popular for measuring brain connectivity, mostly due to its noninvasive nature. In this paper, we are interested in effective connectivity, i.e., the directed influence that one neural system exerts over another, which is often estimated based on resting-state fMRI data \citep{friston2011}.

Statistical approaches to modeling effective connectivity among brain regions include dynamic causal modeling \citep[DCM,][]{friston2003dynamic}, structural equation modeling \citep[SEM,][]{mclntosh1994structural}, Bayesian networks \citep[BNs,][]{li2008dynamic,rajapakse2007learning} and Granger causality modeling via vector autoregressive (VAR) models \citep{granger1969,roebroeck2005mapping}. DCM and SEM are typically used as confirmatory techniques to test pre-defined hypotheses about neural activity \citep{friston2011functional}. Bayesian networks employ directed acyclic graphs, ignoring the high prevalence of reciprocal connections that commonly renders brain connectivity cyclic \citep{friston2011functional}. Granger causality is based on the notion that causes both precede and help predict their effects. This approach infers effective connectivity by estimating coefficients from VAR models.  It is important to note that even though such methods allow inference on directed connections between brain regions, causality between fMRI signals does not translate into causality of the corresponding neuronal activity \citep{wen2013granger}.  

VAR models have been used to estimate whole-brain connectivity networks, where nodes represent brain regions of interest (ROIs) obtained from a parcellation of the brain. Sparsity plays an essential role in the estimation of these models. A common approach is to impose sparsity through an $\ell_1$ penalty on the VAR coefficients \citep{valdes2005,arnold2007}. For fMRI studies with multiple subjects, \cite{gorrostieta2012,gorrostieta2013hierarchical} proposed mixed-effect VAR models that achieve group-level \textcolor{black}{selection, by identifying significant connections between ROIs that are consistent across a group of subjects}. \cite{Chiang2017} considered a supervised setting with subjects belonging to multiple groups and employed spike-and-slab priors to achieve sparsity in the group-level networks. To improve the computational scalability, \cite{kook2021bvar} developed a variational Bayes (VB) approach for the model of \cite{Chiang2017}. See also \cite{MARSSVB} for a recently proposed VB method for an autoregressive state-space model.  \textcolor{black}{While these VAR-based models have significantly advanced the understanding of brain connectivity, they do not account for subject heterogeneity.}

In fMRI studies, it is common to measure subject-level covariates, such as age, sex and behavioral assessment scores, in addition to the imaging data. In recent years, many researchers have focused on the question of how to relate subject-level covariates to the observed imaging data \citep{scheffler2019covariate,guha2021bayesian,kundu2021integrative,zhao2021covariate}. Generally speaking, external covariates can be incorporated into the
estimation of graphs by linking them to either node values (as covariate-adjusted models) or edge strengths (covariate-dependent). However, most of these contributions  in graphical modeling have considered the framework of undirected networks. In this setting, graphs can be expressed as regression models that link the mean values of the network nodes to external covariates, allowing the estimation of a conditional network that reflects dependencies among node variables after adjusting for external covariates \citep{yin2011,li2012}. Time-varying graphs can also be considered a special case of covariate-dependent graphical models, where the network structure is allowed to change smoothly over time \citep{zhou2010,kolar2010time}. More advanced models may allow the external covariates to have non-linear effects on edge strength. These approaches build on regression settings with varying effects, or varying coefficient models (VCM) \citep{cleveland1991,hastie1993}. By modeling regression coefficients as a function of a covariate, these modeling approaches allow greater flexibility than linear regression.

In this paper, we develop an analytical approach for estimating brain connectivity networks that accounts for subject heterogeneity \textcolor{black}{by modeling the effect of covariates on the edge strengths.} 
More specifically, we build upon the VAR framework of \cite{Chiang2017} and \cite{kook2021bvar} to construct a varying-effect vector autoregressive modeling framework that estimates group-specific brain connectivity networks and accounts for the effects of {\color{black}subject-level} covariates on the network edges. We name our method {\it VEVAR} (Varying-Effects Vector AutoRegression). Within this modeling framework, we are able to estimate both a group-level graph structure and  edge strengths as (possibly) non-linear functions of subject-level covariates. We achieve this via Gaussian process priors that capture the edge strengths as smooth functions of covariate values.  We model the covariate effects as a sum of univariate Gaussian processes, which allows for edge-specific covariate effect selection. Additionally, we use variable selection spike-and-slab priors to determine the presence or absence of edges. 
For posterior inference, we address the scalability limitations of the existing models by implementing a variational Bayes approach \citep{blei2017,Titsias2011,dance2023}. 
We use simulated data to compare our method with two-stage frequentist and Bayesian approaches that first estimate the networks and then select the covariates that explain the edge strengths. Our results show that the proposed VEVAR model does well in terms of both group-level edge selection and covariate effect selection. 

Next, we apply our method to resting-state fMRI data from children with a history of traumatic brain injury (TBI) and healthy controls to characterize age and sex effects on neural circuitry. TBI is particularly concerning because it can disrupt the typical course of brain development and lead to cascading effects on health-related quality of life. Furthermore, TBI outcomes are characterized by significant heterogeneity. 
Consequently, statistical approaches that can account for variability related to subject-level covariates can substantially refine the sensitivity and utility of connectivity network modeling. 
Here, in addition to the estimation of the group-level connectivities, we evaluate whether specific group-level edge connectivity strengths are affected by age and sex. A unique feature of our approach is that, unlike other methods, VEVAR allows the estimation of group-level edges as (possibly) non-linear functions of these covariates.    Our results highlight differences in the distribution of parent nodes. They also suggest alteration in the relation of age, with peak edge strength in children with TBI, and differences in effective connectivity strength between males and females. We provide some discussion corroborating our findings.

The rest of the paper is organized as follows.  In Section \ref{Methods and Materials}, we introduce the model and the prior construction and briefly describe the proposed variational approach for inference. We also introduce the simulation design and the empirical data.  In Section \ref{Results}, we assess the performance of our proposed method against competing approaches using simulated data and illustrate our method on resting-state functional MRI data collected on children with a history of traumatic injury and healthy controls. Section \ref{Discussion} provides some interpretation of the results and Section \ref{Conclusion} concludes the paper with a discussion on limitations and future directions.

\section{Methods and Materials} \label{Methods and Materials}
\subsection{Statistical Model} \label{Statistcal Model}
\begin{figure}
    \centering
    \includegraphics[angle=270,width=6.5in]{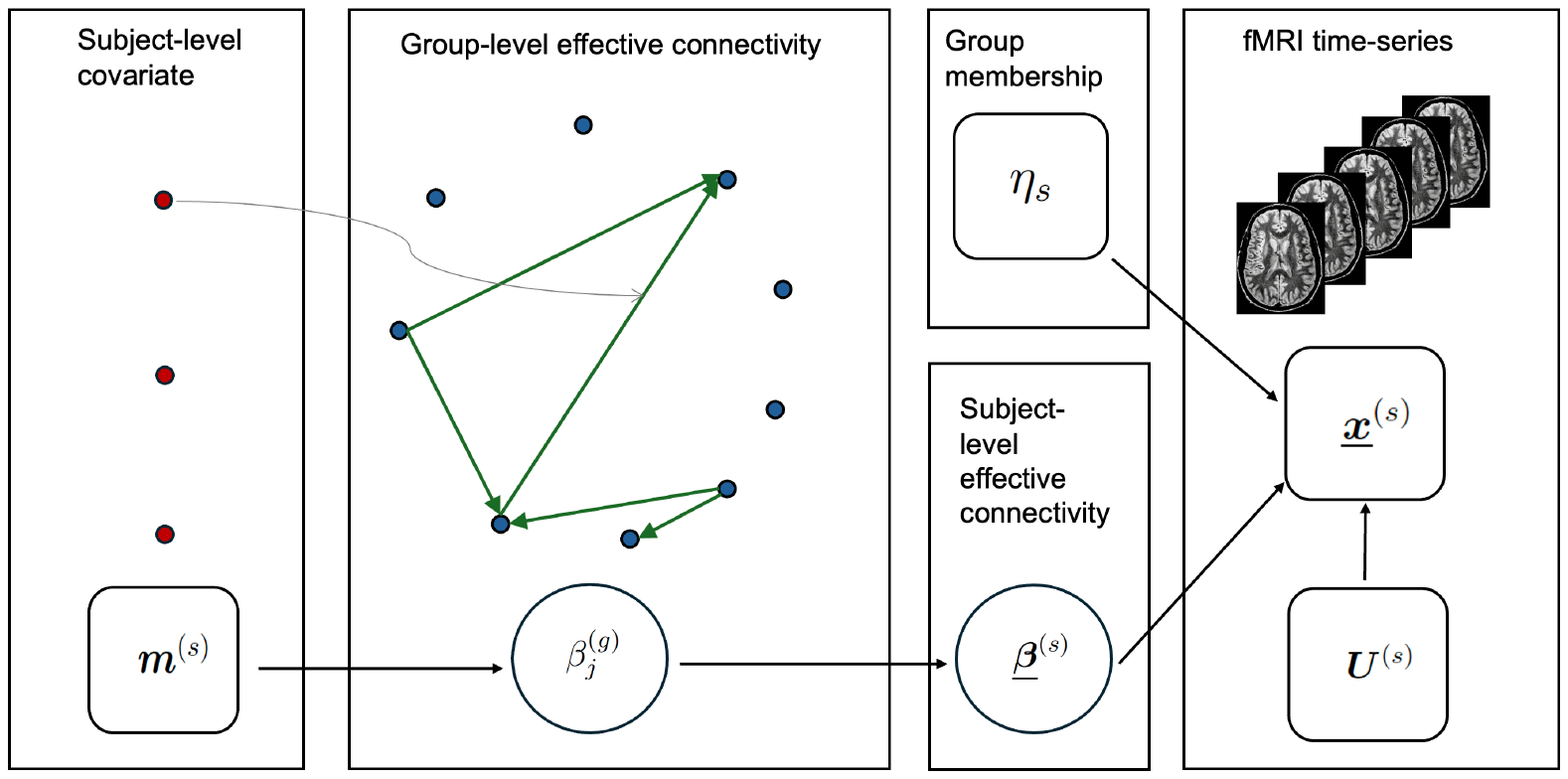}
    \caption{{\color{black}Graphical representation of the proposed model: circular nodes indicate parameters, and square nodes represent observed data. Links between nodes represent direct probabilistic dependence. Some hyperparameters are not shown for clarity. {\color{black}Indices refer to subject, $s=1,\ldots,n$, coefficient, $j=1,\ldots, (RL)R$, and group, $g = 1,\ldots,G$.}}}
    \label{fig:Fig_model}
\end{figure}

We describe the proposed varying-effects VAR model for simultaneous group- and subject-level network estimation and selection of edge-specific covariate effects. \textcolor{black}{A graphical representation of the model is provided in Figure \ref{fig:Fig_model}.}

\subsubsection{Likelihood} \label{Likelihood}
We first introduce the structure of the observed data and the likelihood. In our setting of interest, we observe multivariate time series data on different groups of subjects. Specifically, let ${\bf x}_t^{(s)}=[x^{(s)}_{t,1}, \ldots,x^{(s)}_{t,R}]$ represent the $R$ dimensional vector of observed values for subject $s$ at time $t$, with $s = 1, \ldots, n$, and $t=1, \ldots, T$. We assume that subjects are classified into $G$ groups with $\boldsymbol{\eta} = [\eta_1, \ldots ,\eta_n]$ indicating the group membership for each subject, where $\eta_s=g$ if subject $s$ belongs to group $g=1, \ldots, G$. Furthermore, we assume that an additional set of $P$ fixed covariates $\boldsymbol{m} = [m_{1}, \ldots, m_{P}]$ are observed for each subject.  

Within each group, we model the subject-level time series data via a vector autoregressive (VAR) model that expresses the observed values at time $t$ as a function of the $R$ variables at the previous $L$ lagged time points
\begin{eqnarray}
    \underset{1 \times R}{\underbrace{\vphantom{B^{(s)}_g}\boldsymbol{x}^{(s)}_t}} = 
    \underset{1 \times RL}{\underbrace{\vphantom{B^{(s)}_g}\boldsymbol{u}^{(s)}_t}} \hspace{-2pt} \cdot \hspace{-2pt}
    \underset{RL \times R}{\underbrace{\boldsymbol{B}^{(s)}}}+
    \underset{1 \times R}{\underbrace{\vphantom{B^{(s)}}\boldsymbol{e}^{(s)}_t}},
    \label{eq:model1}
\end{eqnarray}
  where $\boldsymbol{u}^{(s)}_t=[\boldsymbol{x}^{(s)}_{t-1},\boldsymbol{x}^{(s)}_{t-2},\dots,\boldsymbol{x}^{(s)}_{t-L}]$ is the $1 \times RL$ vector of concatenated lagged measurements,
  $\boldsymbol{B}^{(s)}$
  is the $RL \times R$ matrix of subject-specific VAR coefficients, and, given $\eta_s=g$,  $\boldsymbol{e}^{(s)}_t \sim \mathcal{N}(0, \boldsymbol{\Xi}^{(g)})$ with $\boldsymbol{\Xi}^{(g)}=diag(\xi_1^{(g)}, \ldots,\xi_R^{(g)})$ represents independent Gaussian noise. To accommodate all $T$ time points, we can write
  $\boldsymbol{X}^{(s)} = [\boldsymbol{x}_{2}^{(s)}, \ldots, \boldsymbol{x}_{T}^{(s)}]'$, $\boldsymbol{U}^{(s)} = [\boldsymbol{u}_{1}^{(s)}, \ldots , \boldsymbol{u}_{T - L}^{(s)}]'$, and $\boldsymbol{E}^{(s)} = [\boldsymbol{e}_{1}^{(s)}, \ldots, \boldsymbol{e}_{T-1}^{(s)}]'$. We then use the $vec$ operator, which converts a matrix to a single column vector, and can write model \eqref{eq:model1} as
\begin{eqnarray} \label{VAR model}
    \underset{(T - L)R \times 1}{\underbrace{\vphantom{\underline{\beta}_g^{(s)}}\underline{\boldsymbol{x}}^{(s)}}} = 
    \big(\underset{R \times R}{\underbrace{\vphantom{\underline{\beta}_g^{(s)}}I}} \otimes 
    \underset{\hspace{-8pt}(T - L) \times RL}{\underbrace{\vphantom{\underline{\beta}_g^{(s)}}\boldsymbol{U}^{(s)}}}\big)
    \underset{(RL)R \times 1}{\underbrace{\underline{\boldsymbol{\beta}}^{(s)}}} +
    \underset{(T - L)R \times 1}{\underbrace{\vphantom{\underline{\beta}_g^{(s)}}\underline{\boldsymbol{e}}^{(s)}}},
    \label{eq:model2}
\end{eqnarray}
where  $\boldsymbol{\underline{x}}^{(s)}$, $\boldsymbol{\underline{\beta}}^{(s)}$, and $\boldsymbol{\underline{e}}^{(s)}$ are equivalent to $vec(\boldsymbol{X}^{(s)})$, $vec(\boldsymbol{\beta}^{(s)})$, and $vec(\boldsymbol{E}^{(s)})$,  respectively, and $\otimes$ represents the Kronecker product \textcolor{black}{\citep{van2000ubiquitous}}. By writing the model in form \eqref{eq:model2}, we can see that 
\begin{eqnarray}
\underline{\boldsymbol{x}}^{(s)} |\eta_s=g\sim \mathcal{N}((I \otimes \boldsymbol{U}^{(s)})\underline{\boldsymbol{\beta}}^{(s)},\boldsymbol{\Xi}^{(g)} \otimes I),
\end{eqnarray}
which allows us to recognize this as a linear regression problem. \textcolor{black}{The observed values for subject $s$, given the group membership $g$, follow a normal distribution with mean $(I \otimes \boldsymbol{U}^{(s)})\underline{\boldsymbol{\beta}}^{(s)}$ and covariance $\boldsymbol{\Xi}^{(g)} \otimes I$, where $I$ denotes the identity matrix and $\boldsymbol{\Xi}^{(g)}$ is an error covariance matrix} defined as above. The subject-specific coefficients $\underline{\boldsymbol{\beta}}^{(s)} = [\beta_1^{(s)}, \ldots, \beta_{(RL)R}^{(s)}]$ capture the temporal relationship of variables. \textcolor{black}{This model is illustrated in the right portion of Figure \ref{fig:Fig_model}}.

\subsubsection{Varying-effects selection priors}
\cite{Chiang2017} used the VAR setting \eqref{eq:model1} to model fMRI data observed on multiple groups of subjects. In that setting, inference on the VAR regression coefficients allows the estimation of directed networks, where network nodes are brain regions and an edge represents a directional influence 
of one region on another \citep{friston1994}. The authors adopted a hierarchical structure on the $\underline{\boldsymbol{\beta}}^{(s)}$ coefficients by assuming that they are generated from a common group-level coefficient vector $\underline{\boldsymbol{\beta}}^{(g)}$, and imposed spike-and-slab priors on the group-level coefficients. Here, we build upon this supervised hierarchical setting by proposing a novel prior formulation that allows the edge strengths of the group-level graphs to 
vary as functions of the covariate values. We achieve this by imposing nonparametric spike-and-slab priors at the group level that allow edges to be either zero or non-zero functions of the covariate values. Additionally, we employ univariate spike-and-slab priors to select covariates influencing individual edge strengths, for each subject group. Our novel prior formulation allows for non-linear effects of covariates on the edge strengths of the inferred networks and simultaneous selection of the relevant effects.

As in \cite{Chiang2017}, we assume that the VAR coefficients for each subject, $\beta^{(s)}_{j}$, are noisy realizations of a group-level network. However, at the group level, we model the edge strengths as functions of the subject-specific covariate values, $\boldsymbol{m}$.
 We write
\begin{eqnarray}
  \beta^{(s)}_{j} \mid \boldsymbol{m} \sim & \mathcal{N}(\beta_j^{(g)}(\boldsymbol{m}),\sigma_j^{(g)}),
  \label{eq:subject coefs}
 \end{eqnarray}
for all subjects $s$ such that $\eta_s = g$, with  $j=1,\ldots, (RL)R$ and $g = 1,\ldots,G$.
We place the priors $\sigma_j^{(g)} = \sigma_0^{(g)} \sim IG(a_0^{(g)},b_0^{(g)})$ for $\delta_j^{(g)}=0$ and $\sigma_j^{(g)} = \sigma_1^{(g)} \sim IG(a_1^{(g)},b_1^{(g)})$ for $\delta_j^{(g)}=1$. We seek to estimate a network for each of the $G$ groups, that is, to find the non-zero group-level edges. We achieve this by proposing a novel discrete nonparametric spike-and-slab prior that imposes sparsity on the networks at the group level while modeling the non-zero edges as smooth functions of the covariate values:
\begin{eqnarray}
    \beta^{(g)}_j(\boldsymbol{m}) =  \delta_j^{(g)} f_j^{(g)}(\boldsymbol{m}) + (1 - \delta_j^{(g)})\delta_0(\boldsymbol{m}),
\label{eq:Group Spike and Slab}
\end{eqnarray}
with $\delta_j^{(g)} \sim Bernoulli(\pi_\delta)$, where the functional notation $\beta^{(g)}_j(\boldsymbol{m})$ represents possible values of the VAR coefficient as a function of the covariate values $\boldsymbol{m}$, the ``spike" $\delta_0$ represents a Dirac delta function, and the notation $\delta_j^{(g)} = 0$ implies that the group-level coefficient for group $g$ is zero for all possible values of $\boldsymbol{m}$. Spike-and-slab priors typically include a normal distribution as the ``slab" portion \citep{GeorgeMcCulluch,vannucci2021discrete}. We innovate on this framework, to instead allow the strengths of the selected nonzero edges to depend on covariate values in a non-linear manner, as described below. A nice result of this modeling choice is that we can obtain a {\it set} of directed edges for each group, i.e., a form of interpretable functional group networks where the strengths of the included edges vary as smooth functions of the covariate values. Our nonparametric ``slab" portion of the mixture prior \eqref{eq:Group Spike and Slab} is defined as follows. We first incorporate the covariates as
\begin{eqnarray}
    f_j^{(g)}(\boldsymbol{m}) =  \mu_j^{(g)} + \sum_{p = 1}^P w_{j,p}^{(g)}  \phi_{j,p}^{(g)}(m_p),
    \label{eq:group function}
\end{eqnarray}
with $\mu_j^{(g)}\sim\mathcal{N}(0,\sigma^2_\mu)$ a baseline value reflecting edge strength not driven by covariate effects,  $\phi_{j,p}^{(g)}(m_p)$ a univariate function dependent on the $p$th covariate, and $w_{j,p}^{(g)}$ the corresponding weight. We then model the functions $\phi_{j,p}^{(g)}(m_p)$ via Gaussian process priors \citep{williams2006gaussian} as
\begin{eqnarray}
  \phi_{j,p}^{(g)}  \equiv 
         GP(\mu(\cdot),K(\cdot,\cdot)),  \label{eq:GP} 
\end{eqnarray}
with $\mu(\cdot) = 0$ and covariance function 
$K(m_{p,k},m_{p,k'}) = \text{cov}(f(m_{p,k}),f(m_{p,k'}))$,
which allows us to model non-linear effects of the covariates on edge strength. 
Finally, we impose a parametric discrete spike-and-slab prior on the coefficients in \eqref{eq:group function} as 
\begin{eqnarray}
w_{j,p}^{(g)} \sim  \pi_\phi \mathcal{N}(0,\sigma^2_w) + (1 - \pi_\phi) \delta_0, \label{eq:spike and slab}
\end{eqnarray}
where $\pi_\phi$ is the prior probability that the coefficient is non-zero, i.e., the corresponding covariate is determined to be important in estimating the edge strength function, $\beta_j^{(g)}(\boldsymbol{m})$.
For each edge $j$ in group $g$, formulation \eqref{eq:group function} employs $P$ non-linear functions, $\phi_{j,p}^{(g)}(m_p)$, one for each covariate. Thus, using the spike-and-slab prior \eqref{eq:spike and slab} on the individual $w_{j,p}^{(g)}$'s allows us to select possibly different covariates, with non-linear effects, for each group-level edge. We complete our model by assuming priors $\xi_r^{(g)} \sim IG(a_\xi,b_\xi)$, for $r=1,\ldots, R$, on the diagonal elements of $\boldsymbol{\Xi}^{(g)}$.

\begin{figure}
\centering
    \includegraphics[width = 6.0in]{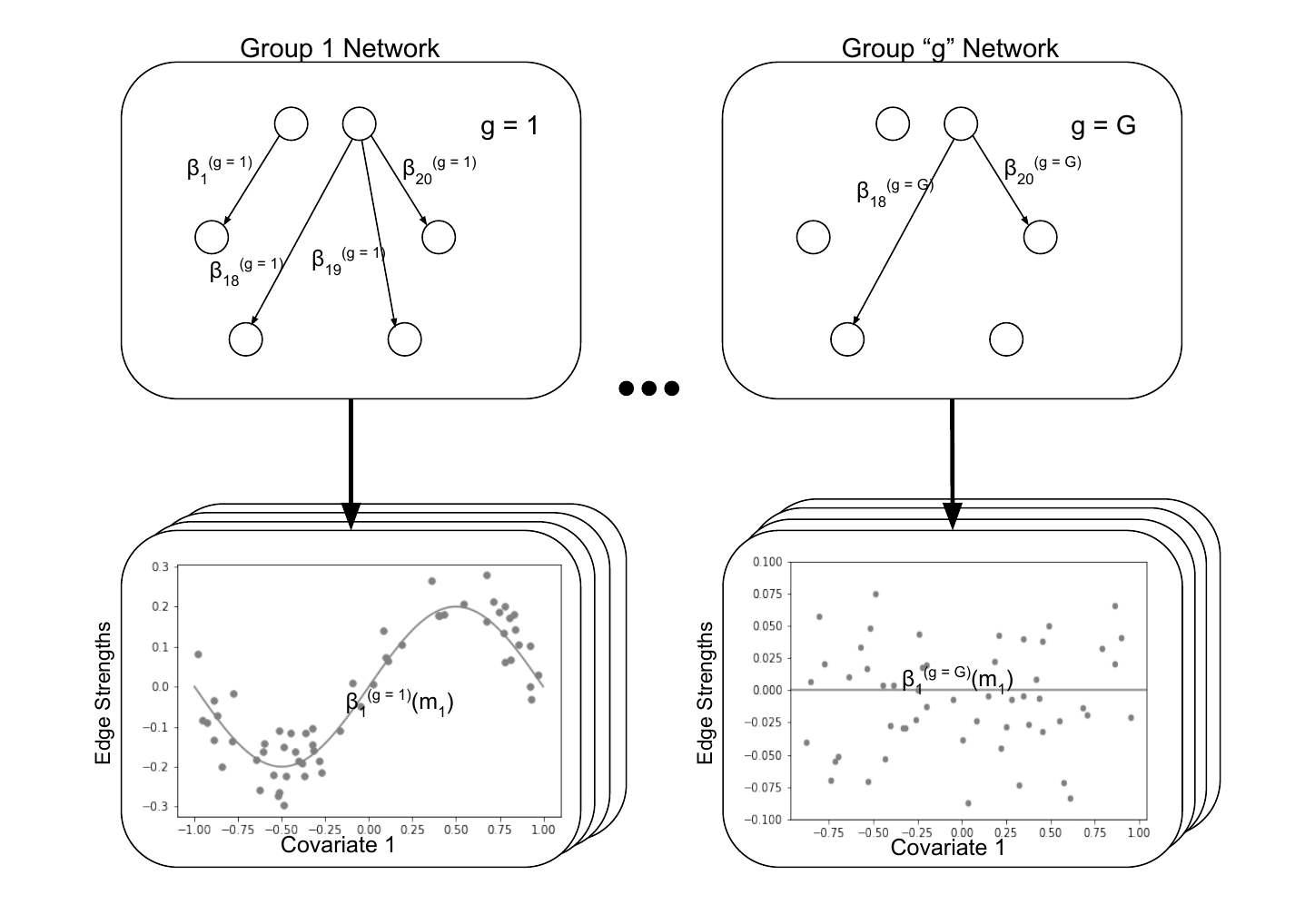}
        \caption{An example diagram of the proposed varying-effects selection prior, with an illustration of group-level edges, subject-level coefficients, and group-level coefficient functions. The top row shows group-level networks. The lower row shows the underlying group level function, $\beta^{(g)}_j(\boldsymbol{m})$, as a function of a single covariate, $m_1$, for a single edge, $j=1$. Subject-level edge strengths, $\beta^{(s)}_1(m_1)$, are shown as points following the shape of the function $\beta^{(g)}_1(m_1)$.}
         \label{fig:diagram}
\end{figure}

The proposed modeling framework described here represents a novel approach to inference of covariate effects on edge strength within a VAR model, that leads to the estimation of sparse networks at the group level and the selection of key covariates that influence individual edge strengths. Figure \ref{fig:diagram} shows an illustration of our proposed varying-effect nonparametric selection prior. The top row shows group-level networks, which will be estimated using the spike-and-slab prior \eqref{eq:Group Spike and Slab}. The lower row shows the underlying group-level function, $\beta^{(g)}_j(\boldsymbol{m})$, modeled via the nonparametric slab \eqref{eq:group function} incorporating the covariates, as a function of a single covariate, $m_1$, and for a single edge, $j=1$. Subject-level edge strengths, $\beta^{(s)}_1(m_1)$, for all $s$ such that $\eta_s=g$, are shown as points following the shape of the function $\beta^{(g)}_1(m_1)$. Notice that the edge corresponding to $\beta_{1}^{(g)}$ is present in the group-level network for $g=1$ but not for $g = G$. 

The GP construction in \eqref{eq:GP} requires selecting a kernel function. Here we used a squared exponential (SE) of the form $K(x,x') = \sigma^2\text{exp}(-\frac{(x-x')^2}{2l^2})$. We found a length scale fixed at $l=0.5$ to be sufficiently flexible for our application. In addition, we found this kernel to work for both continuous and categorical covariates, as shown by our results in the simulation study presented below. Our methodology is general and can accommodate any valid kernel function. Moreover, a common kernel does not need to be used across all covariates or across all edges. Additional flexibility could be gained by placing a prior on any of the hyperparameters of the kernel.

\subsubsection{Variational algorithm for scalable inference} \label{Model Estimation}
One challenge in network estimation is the large number of parameters that need to be estimated. This is particularly challenging in Bayesian estimation, as Bayesian methods frequently rely on Markov chain Monte Carlo (MCMC) methods \citep{gelman2013bayesian} to sample from the posterior. In our model formulation, complexity and computational costs are compounded by the consideration of the covariates and the large number of inclusion indicators we need to estimate in the nonparametric spike-and-slab construction. 

To allow the model to scale up to large data sizes, in particular those encountered in our application setting, we implement a variational approximation method for posterior inference which dramatically improves computational time. Variational schemes for linear models that use spike-and-slab priors have been previously described \citep{carbonetto,Titsias2011,kook2021bvar}.  Variational inference (VI) aims at finding an approximation of the posterior by using optimization methods. It works by specifying a family of approximate distributions, $\mathcal{Q}$, which are densities over model parameters and latent variables that depend on free parameters $\boldsymbol{\Theta}$, and then seeks to find the values of $\boldsymbol{\Theta}$ that minimize the Kullback-Leibler (KL) divergence between the approximate distribution and the true posterior. Let $\mathbf{Z}$ indicate the
set of model parameters and latent variables. As discussed in \cite{blei2017}, minimizing the KL divergence is equivalent to maximizing the Evidence Lower BOund (ELBO), defined as
\begin{align}
    \text{ELBO} = \EX_{\Theta} [\text{log } p(\mathbf{Z} , Y)] - \EX_{\Theta} [\text{log } q(\textbf{Z})],
    \label{eq:ELBO_simple}
\end{align}
with $p(\mathbf{Z},Y)$ the joint distribution of ${\bf Z}$ and the data, and $q(\mathbf{Z})$ the approximate distribution. The complexity of the optimization procedure is determined by the complexity of the variational distribution. A common family of approximate distributions is the mean field approximation, which assumes that the approximate distribution factorizes over some partition of the parameters and latent variables as 
\begin{equation}
q(\mathbf{Z}) = \prod_k q(Z_k).
\end{equation}
The exact parametric form of each $q(Z_k)$ is selected based on whether $Z_k$ is continuous or discrete, and may exploit conditional conjugate distributions to allow for simpler derivation of the optimization steps.  

In our framework, we adopt a mean field approach and introduce a family of approximate distributions $Q(\boldsymbol{\Theta})$.
We follow the work of \cite{Titsias2011} by reparameterizing the spike-and-slab prior in \eqref{eq:spike and slab} through the introduction of new variables $\tilde{w}_{j,p}^{(g)} \sim \mathcal{N}(0,\sigma_w^2)$ and $s_{j,p} \sim \pi_\phi^{s_{j,p}} (1 - \pi_\phi)^{1 - s_{j,p}}$. This allows us to rewrite $w_{j,p}^{(g)}$ as $ w_{j,p}^{(g)} =\tilde{w}_{j,p}^{(g)}s_{j,p}^{(g)}$. We then find $q(\tilde{w}_{j,p}^{(g)} \mid s_{j,p})$ and $q(s_{j,p}^{(g)})$ and model $\{\tilde{w}_{j,p}^{(g)},s_{j,p}^{(g)}\}$ jointly, according to the mean field approach. The remaining variational distributions are proposed as:

\begin{eqnarray}
    q(\mu_j^{(g)}) &\sim& \mathcal{N}(u_j^{(g)},v_j^{(g)}) \label{eq:q(mu)} \nonumber
    \\
    q(\delta_j^{(g)}) &\sim&  \text{Bernoulli}(\gamma_{(\delta),j}^{(g)}) \label{eq:q(gamma_delta)} 
    \nonumber
    \\
    q(\underline{\boldsymbol{\beta}}^{(s)}
    \mid \underline{\boldsymbol{\beta}}^{(g)}(\boldsymbol{m}^{(s)}), \sigma_j) &\sim & \mathcal{N}(\boldsymbol{u}^{(s)}_\beta,\boldsymbol{S}^{(s)}) 
    \nonumber 
    \\
    q(\xi_r^{(g)} \mid a_\xi^{(g)},b_\xi^{(g)}) &\sim& IG(z_{1r}^{(g)},z_{2r}^{(g)}) 
    \label{eq:variational distributions}
    \nonumber
    \\
    q(\sigma_0^{(g)} \mid a_0^{(g)},b_0^{(g)}) &\sim& IG(a_0^{(g)},b_0^{(g)}) 
    \label{eq:q(sigma)}
    \\
    q(\sigma_1^{(g)} \mid a_1^{(g)},b_1^{(g)}) &\sim& IG(a_1^{(g)},b_1^{(g)}) 
    \label{eq:q(sigma)}
    \nonumber
    \\
    q(\phi^{(g)}_{j,p}(\boldsymbol{m})) &\sim & \mathcal{N}(\tilde{\boldsymbol{\mu}}_{j,g}^{\phi},\tilde{\textbf{K}}_{j,g}^{\phi})
    \label{eq:q(phi)}
    \nonumber
    \\
    q(\tilde{w}_{j,p}^{(g)} \mid s_{j,p}^{(g)}) &\sim & \mathcal{N}(\omega_{j,p}^{(g)},\tilde{\sigma}_{j,p}^{(g)})
    \label{eq:q(omega)}
    \nonumber
    \\
    q(s_{j,p}^{(g)}) &\sim&  \text{Bernoulli}(\gamma_{(\phi),j,p}^{(g)}),
    \label{eq:q(s)}
    \nonumber
\end{eqnarray}
\noindent where $\underline{\boldsymbol{\beta}}^{(g)}(\boldsymbol{m}) = [\beta_j^{(g)}(\boldsymbol{m}), \ldots , \beta_{(RL)R}^{(g)}(\boldsymbol{m})]$ is the collection of realizations of the group-level function at subject $s$'s covariate measurements,
and $\tilde{\textbf{K}}_{j,g}^{\phi}$ is the covariance matrix corresponding to the kernel $K(\boldsymbol{m}_p^{(g)},\boldsymbol{m}_p^{(g)})$ where $\boldsymbol{m}_p^{(g)} = [m_p^{(s = 1)}, \ldots , m_p^{(s = N)}]$ for $\eta_s = g$. 
Using a mean field family, and the reparameterizations discussed above, we write the full approximate distribution for all variational parameters as a product over the approximate distributions:
\begin{eqnarray*}
Q(\boldsymbol{\Theta})& =& \prod_{g = 1}^G  \bigg [ \prod_{j = 1}^{R \times R \times L} \Big\{ 
q\big(\mu_j^{(g)}\big) 
q\big(\delta_j^{(g)}\big)
  \prod_{p = 1}^P 
q\big(\phi^{(g)}_{j,p}\big)
q\big( \tilde{w}_{j,p}^{(g)} \mid s_{j,p}^{(g)} \big)  
q\big(s_{j,p}^{(g)}\big) 
\Big\} \\
&\times&\prod_{r = 1}^R 
q(\xi_r^{(g)})
\bigg] \prod_{s = 1}^N q\big(\beta^{(s)}_j
\big)q\big(\sigma_1^{(g)}\big)
q\big(\sigma_0^{(g)}\big), 
\end{eqnarray*}
\noindent where $\boldsymbol{\Theta}$ represents the parameters to be optimized. 
As common with variational inference, these are updated via coordinate ascent variational inference (CAVI). This algorithm is repeated until the ELBO has converged or has changed by some minimum threshold. Updates of the parameters and the parameter blocks are outlined in the Supplementary Material.
Posterior inference based the VI algorithm results in estimates of the group-level networks, together with estimates of the edge strengths at the group- and subject-level, and simultaneous selection of the covariates that are relevant to the edge strengths. After convergence of the VI algorithm, we classify edge $j$ as being present in group $g$ if the estimated $\gamma_{(\delta),j}^{(g)}$ is greater than 0.5, and select covariate $p$ as affecting  the $j^{th}$ edge strength in group $g$ if the estimated $\gamma_{(\phi),j,p}^{(g)}$ is greater than 0.5. We infer the non-zero group-level edge strength functions by the values of $u_j^{(g)} + \sum_{p = 1}^P \gamma^{(g)}_{(\phi),j,p}\omega^{(g)}_{j,p}\tilde{\boldsymbol{\mu}}_{j,g}^{\phi}$ in \eqref{eq:variational distributions} and the subject-level edge strengths by the values of $\boldsymbol{u}_\beta^{(s)}$ in \eqref{eq:variational distributions}.
We note that variational approaches are only suitable for point estimation and do not allow the assessment of uncertainty about the estimates. Additionally, in situations with correlated covariates, performance can be sensitive to initializations and to the order that variables are updated, possibly resulting in poor selection performance. \cite{ray2021} proposed a prioritized updating scheme, where the importance of the variable is determined using a preliminary estimator. Here we obtained the preliminary estimator as the average of the output from several cold starts, where each covariate is afforded the opportunity to be the first and also the last covariate proposed. 

\subsection{Simulation experiment} \label{Simulation}
We use simulated data to test the performance of the proposed model and compare results to competing approaches. 

In our simulation, we considered two groups of subjects with sample sizes 30 and 60, respectively, to assess the robustness of the model to an unbalanced sample size setting. For each subject, we generated the time series data $\boldsymbol{x}_{t}^{(s)}$ {\color{black}to represent realistic fMRI signal across ROIs over time.  This data was drawn from a multivariate normal distribution} $\mathcal{N}(\boldsymbol{u}_{t}^{(s)} \boldsymbol{B}^{(s)},\sigma^2I_R)$, for $t = 2, \ldots, 200$, with noise variance $\sigma^2=0.5$. {\color{black}We selected $R=100$ ROIs to mirror the level of granularity commonly observed in brain network studies. This choice ensures that our simulation represents the complex connectivity between different brain regions.} The initial time point  $\boldsymbol{x}_{1}^{(s)}$ was randomly drawn from $\mathcal{N}(\boldsymbol{0},0.25I_R)$. 
We simulated the VAR coefficients $\boldsymbol{B}^{(s)}$, {\color{black}representing the connectivity strength and direction between different ROIs,} in a manner that allowed us to test a range of possible functions and covariate associations. {\color{black}The time lag value was set to $L=1$, ensuring the simulation of direct, immediate connections typical in brain networks. We checked to make sure that the process generated only stationary time series for each subject, and repeated the data generating process if not.} We considered six random subject-level covariates, $\textbf{m}=(m_1,\ldots,m_6)$, with the first five drawn from a uniform$(-1,1)$, and the sixth one from a Bernoulli$(0.5)$. {\color{black}These covariates affect the edge strengths within the brain connectivity network, inducing individual differences in neural connections. }
Next, we generated each element of the subject level vector of coefficients $\boldsymbol{B}^{(s)}$ from a $\mathcal{N}(f(\textbf{m}),0.08)$, with $f(\cdot)$ the generating function of covariates on the edge strengths defined by setting each element $B_{j,j'}^{(s)}$, with $j = 1,\ldots,RL$ and $j' = 1,\ldots R$, as follows:
\begin{align}
\nonumber
    if \mid j - j' \mid = 0 &, f(\cdot) = 0.15 \text{~~(i.e., constant}) \\\nonumber
    if \mid j - j' \mid = 1 &, f(\cdot) = m_p \cdot 0.25 \\\nonumber
    if \mid j - j' \mid = 2 &, f(\cdot) = (1.0 + m_p)^{0.40} \\\label{eq:functions}
    if \mid i - j' \mid = 3 &, f(\cdot) = (m_p)^{2} \\\nonumber
    if \mid j - j' \mid = 4 &, f(\cdot) = 0.20 \cdot\text{sine}(\pi m_p) \\\nonumber
    if \mid j - j' \mid = 5 &, f(\cdot) = 0.20 \cdot m_6 \\\nonumber
    if \mid j - j' \mid = 6 &, f(\cdot) = 0.3 m_{p_1} - 0.3 m_{p_2} \\\nonumber
    if \mid j - j' \mid = 7 &, f(\cdot) = (1.0 + m_p)^{0.70},\nonumber
\end{align}

\noindent where $m_p$ is randomly selected from the first 5 covariates, and $m_{p_1}$ and $m_{p_2}$ refer to two randomly selected covariates from the first 5 covariates. Note that $m_p, m_{p_1}$, and $m_{p_2}$ are randomly selected separately for each $B_{j,j'}^{(g)}$. Additionally, $f(\cdot)$ was rescaled to be within $[-0.2,0.2]$ in the case where $\mid j - j' \mid$ is equivalent to one of $[2,3,4]$, and within $[-0.15,0.35]$ when $\mid j - j' \mid = 7$, to explore scenarios with both a non-zero mean and covariate effect. Additionally, to more easily sample stationary time series, each function was multiplied by -1 with probability 0.5. 
Finally, to allow differences between the two sample groups, while remaining mostly similar, each $f^{(g)}(\cdot)$ outlined above was set to 0 with probability 0.2, resulting in some group-level edges being present in one group but not the other. \textcolor{black}{By not restricting our simulation to predefined patterns, we can capture the diverse and complex network structure observed in the brain. Furthermore, the magnitude and variance of the edge strengths in our simulation are designed to replicate the estimated values in previous fMRI studies \citep{kook2021bvar}.}

\subsection{Observational Study on Traumatic Brain Injury}
\label{sec:fMRI}
We analyzed data from an fMRI study on children with a history of traumatic brain injury (TBI) following a vehicle collision and healthy controls. Subjects were recruited from the Emergency Department or Level 1 Pediatric Trauma Center at the Children’s Memorial Hermann Hospital, University of Texas Health Science Center at Houston (UTHealth), between September 2011 and August 2015 \citep{watson2019graph,ewing2019post,kook2021bvar}. 

\subsubsection{Participants}
Participants were included in the study if they met the following criteria: 1) injured in a vehicle accident between 8 and 15 years of age; 2) proficiency in English or Spanish; 3) residing within a 125 mile catchment radius; 4) no prior history of major neuropsychiatric disorder (intellectual deficiency or low-functioning autism spectrum disorder) that would complicate assessment of the impact of injury on brain outcomes; 5) no metabolic, endocrine, or systemic health problems (e.g., hypertension); 6) no prior medically-attended TBI; and 7) no habitual use of steroids, tobacco, or alcohol. The latter four criteria were assessed during screening using a brief parent interview.  
A quality control evaluation of all scans resulted in 70 TBI samples being selected for analysis. Additionally, 50 subjects with no history of trauma were measured as part of a healthy control (HC) group. 
Subjects were excluded for excessive motion or scanner error (e.g., operator error, crack in scanner head coil). 
Written informed consent was obtained from each child’s guardian and written assent was obtained from all children in accordance with Institutional Review Board guidelines. \textcolor{black}{The demographic characteristics of the participants in the study are shown in Table \ref{tab:demographics}.}

\begin{table}
\centering
\begin{tabular}{|l|c|c|}
\hline
& TBI (n=70) & HC (n=50) \\
\hline
Age (years)& 12.54 (8.08-16.0)& 12.13 (8.08-16.0) \\
Sex (M, \%) & 45 (64\%) & 30 (60\%) \\  
\hline
\end{tabular}
\caption{\textcolor{black}{Demographic information for participants. Age is summarized as mean (range), and sex is summarized as number (percentage).}}
\label{tab:demographics}
\end{table}

\subsubsection{Data pre-processing}
The fMRI data was preprocessed using SPM12 from the Wellcome Trust Center for Neuroimaging (http://www.fil.ion.ucl.ac.uk/spm/). The preprocessing steps involved correcting motion through realignment, correcting slice timing, separating gray matter, white matter, and CSF, registering the data to the subject's T1-weighted MPRAGE image and a standard space using the ICBM space template, and smoothing with an 8mm full-width half maximum (FWHM) Gaussian kernel. The data were then screened using the Artifact Detection Tools (ART) toolbox \citep{whitfield2011artifact} to eliminate volumes with excessive motion. Participants with more than 15\% of their volumes affected by motion outliers were excluded from the analysis. Finally, a 3D parcellation was performed using the MarsBaR toolbox in SPM 12 and the automated anatomical labelling (AAL) brain atlas, resulting in 90 ROIs excluding regions associated with the cerebellum.

\subsubsection{Previous findings}
In \cite{kook2021bvar} and \cite{Vaughn}, Bayesian vector-autoregressive models were employed to identify unique effective connectivity patterns within the default mode network in children with TBI relative to healthy controls. 
The connectivity patterns differed according to the severity of TBI and showed specific directional relations with symptom profiles. Fewer post-concussion and anxiety symptoms were associated with stronger regional effective orbitofrontal to posterior cingulate cortex connectivity for mild TBI, whereas weaker connectivity was associated with better outcomes for more severe TBI \citep{Vaughn}. These findings were similar to prior reports in adults with TBI showing different relations of frontal lobe connectivity with outcomes after mild TBI versus more severe TBI \citep{zhou2012default, wu2015intrinsic}.

\section{Results}\label{Results}
\subsection{Simulation results}
We fit our proposed model by setting parameters $\pi_\phi = \pi_\delta = 0.1$ for the selection priors on covariate effects and group-level edges, respectively. We also set $\sigma_w = 1$ as the variance of the slab in \eqref{eq:spike and slab} and $\sigma_\mu = 1$ as the variance of the baseline in \eqref{eq:group function}. For the priors on the variance of the subject-level edge strengths, i.e., $\sigma_0^{(g)} \sim IG(a_0^{(g)},b_0^{(g)})$ and $\sigma_1^{(g)} \sim IG(a_1^{(g)},b_1^{(g)})$, we imposed vague priors by setting $a_0 = a_1 = 2$ and $b_0 = b_1 = 1$. Similarly, for the prior on the variance of the time series data $\boldsymbol{x}$, i.e. $\xi_r^{(g)} \sim IG(a_\xi,b_\xi)$, we set $a_\xi = 2, b_\xi = 1$. For the kernel of the Gaussian process, we set length-scale $l = 0.5$ and output variance $\sigma^2 = 1$. A sensitivity analysis of the parameter choices is provided in Section \ref{sec:sens} below.
\textcolor{black}{We executed our simulations on a MacBook Pro with an M1 chip. The running time ranged from 12 to 16 hours, according to the convergence speed of the VI algorithm for the different replicated datasets.}

\subsubsection{Comparative performance}
Table \ref{table:NetworkCovResults} provides results for group-level edge selection and for covariate effect selection, averaged over 25 replicated datasets, in terms of true positive rate (TPR), false positive rate (FPR), Matthew's correlation coefficient (MCC), F1 score, and accuracy \textcolor{black}{(Acc). Specifically, the metrics are defined as
\begin{eqnarray*}
    \text{TPR} &=& \frac{\text{TP}}{\text{TP} + \text{FN}},\\
    \text{FPR} &=& \frac{\text{FP}}{\text{FP} + \text{TN}},\\
    \text{MCC} &=& \frac{\text{TP}\times\text{TN} - \text{FP}\times\text{FN}}{(\text{TP} + \text{FP})(\text{TP} + \text{FN})(\text{TN} + \text{FP})(\text{TN} + \text{FN})},\\
    \text{F}_1 &=& \frac{2\text{TP}}{2\text{TP} + \text{FP} + \text{FN}},\\
    \text{Acc} &=& \frac{\text{TP}+\text{TN}}{\text{TP}+ \text{TN} + \text{FP} + \text{FN}},
\end{eqnarray*}
where TP, TN, FP, and FN denote the number of true positives, true negatives, false positives and false negatives, respectively.}
In the same table, we report results from alternative approaches. Since we are not aware of other methods that achieve simultaneous group-level edge selection and covariate effect selection, we considered two-stage approaches that, at the first stage, estimate the networks and at the second stage select the covariates that explain the edge strengths. For one approach, at the first stage, i.e., group-level edge selection, we considered a traditional Granger Causality (GC) model. \textcolor{black}{Granger Causality \citep{granger1969investigating} is a statistical hypothesis test to assess whether one time series can predict another time series. Here,} it estimates subject-level VAR coefficients via ordinary least squares and then performs group-level inference through one-sample t-tests. Non‐zero group-level edges were identified by thresholding $p$‐values with false discovery rate control at the 0.05 level. At the second stage, i.e., covariate effect selection, we regressed the subject-level strengths estimated via the GC method on the covariates and performed selection using LASSO \citep{LASSO} (GC-LASSO), which identifies linear covariate effects, and plsmselect \citep{plsmselect} (GC-plsmselect), \textcolor{black}{which fits Generalized Additive Models (GAMs) with flexible penalties, allowing for linear and smooth covariate effects.} 
We also considered a second two-stage approach, obtained by considering at the first stage model \eqref{eq:group function} with the GP component removed and only an intercept term included. The resulting model, which we call VEVAR-S1, estimates subject- and group-level networks without accounting for covariate effects. At the second stage, we regressed the estimated subject-level coefficients on the covariates and performed variable selection via the model of \cite{carbonetto} (VEVAR-S1-SS), \textcolor{black}{which uses mixture priors on the regression coefficients to perform variable selection}. Given the relatively low FPRs for edge selection using GC and VEVAR-S1, we performed selection at the second stage by using all estimated edges rather than only the selected ones.

\begin{table}
\begin{center}
\begin{tabular}{|l|c|c|c|c|c||c|c|c|c|c|}
      \multicolumn{11}{c}{Edge Selection}\\
      \hline 
      & \multicolumn{5}{|c||}{Group 1} & \multicolumn{5}{|c|}{Group 2} \\
      \hline
     & TPR & FPR & MCC & F1 & ACC & TPR & FPR & MCC & F1 & ACC \\
     \hline
    GC & .373 & .060 & .339 & .407 & .874 & .456 & .064 & .400 & .467 & .881 \\        
    VEVAR-S1 & .560 &  \textbf{.003}   & .711 & .707 & .947 & .633 & \textbf{.018}    & .692 & .716 & .942 \\
    VEVAR & \textbf{.935} & .029  & \textbf{.852} & \textbf{.863} & \textbf{.962} & \textbf{.999} & .039 & \textbf{.859} & \textbf{.868} & \textbf{.965}  \\        
\hline 
      \multicolumn{11}{c}{Covariate Effect Selection}\\
      \hline
      & \multicolumn{5}{|c||}{Group 1} & \multicolumn{5}{|c|}{Group 2} \\
      \hline 
      & TPR & FPR & MCC & F1 & Acc & TPR & FPR & MCC & F1 & Acc \\
      \hline 
     GC-LASSO &  .275 & .030 & .190 & .204 & .950 & .276 & .022 & .220 & .236 & .964\\
     GC-plsmselect  &  .364 & .127 & .099 & .098 & .863 & .359 & .105 & .115 & .112 & .884\\
     VEVAR-S1-SS &  .693 & .020 & .532 & .527 & .974 & .780 & .025 & .546 & .528 & .971\\ 
     VEVAR & \textbf{.879} & \textbf{.000} & \textbf{.936} & \textbf{.935} & \textbf{.998} & \textbf{.998} & \textbf{.000} & \textbf{.999} & \textbf{.999} & \textbf{.999}\\
\hline 
\end{tabular}
\caption{\textcolor{black}{{\bf Simulation study:} Results of competing methods and VEVAR for group-level edge selection and covariate effect selection, evaluated using true positive rate (TPR), false positive rate (FPR), Matthew's correlation coefficient (MCC), F1 score, and accuracy (Acc). GC indicates Granger causality, and VEVAR-S1 refers to the VEVAR model without covariate effects. GC-LASSO employs LASSO to select covariates affecting edge strengths estimated by GC, while GC-plsmselect utilizes the method described in \cite{plsmselect}. VEVAR-S1-SS applies the spike-and-slab approach from \cite{carbonetto} for selecting covariates impacting edge strengths in VEVAR-S1. Bold values denote the best performance for each metric.}}
\label{table:NetworkCovResults}
\end{center}
\end{table}

Results from Table \ref{table:NetworkCovResults} show that our proposed VEVAR model does well in terms of all performance measures, for both group-level edge selection and covariate effect selection. For edge selection, VEVAR does considerably better than the other methods in all metrics, 
with only slightly higher false positive rates compared to VEVAR-S1. This is because the inclusion of covariate effects introduces more variability while improving the edge selection. For covariate effect selection, we see again that VEVAR performs the best across all metrics. 

\begin{figure}
\centering
    \includegraphics[width = 6.5in]{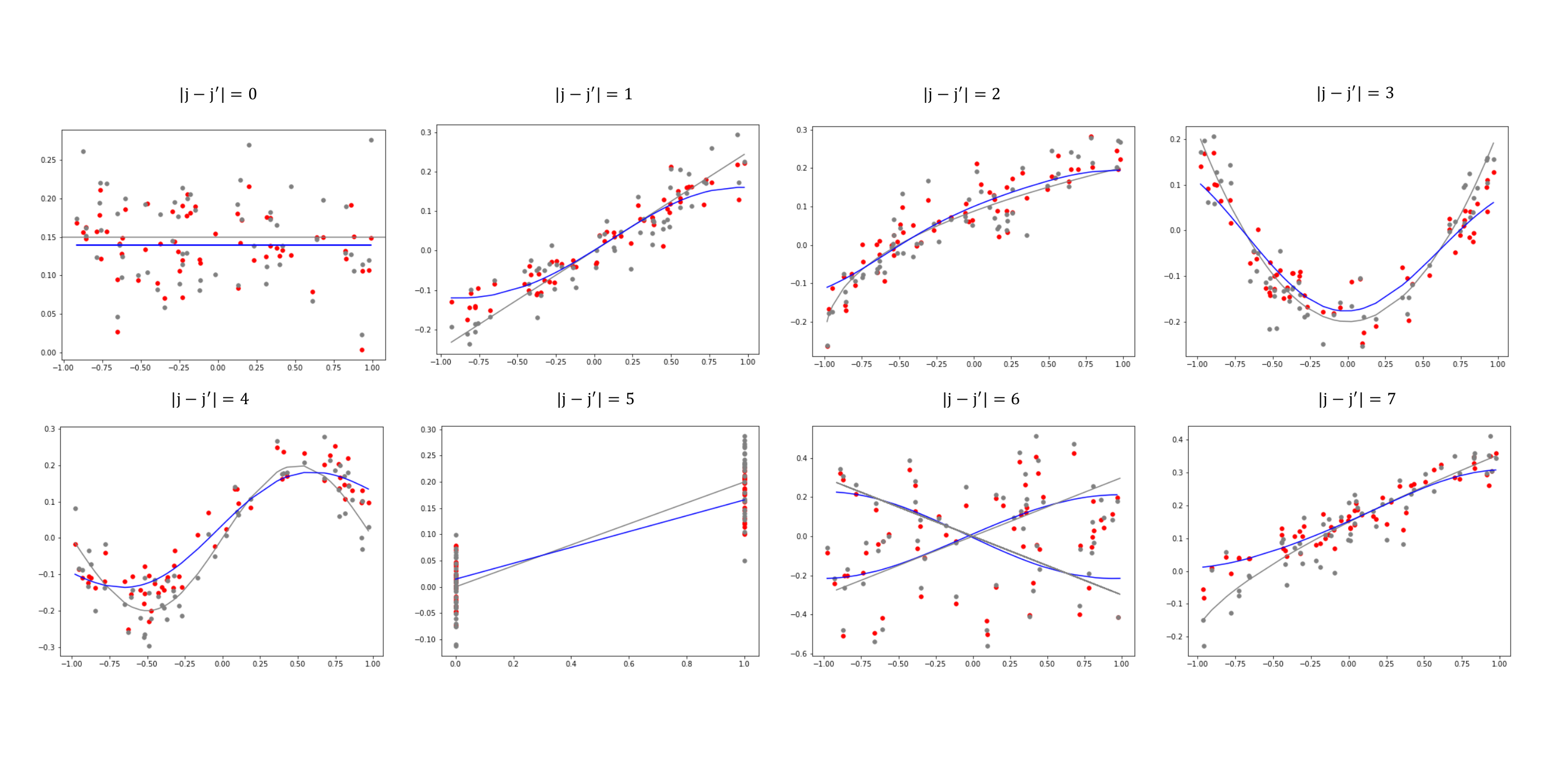}
    \caption{{\bf Simulation study:} True and estimated edge strengths and group-level functions. Gray points represent simulated subject-level edge strengths and gray lines represent the true group-level functions. Blue points represent estimated subject-level edge strengths and blue lines indicate the estimated group-level functions. }
    \label{fig:EstimatedSim}
\end{figure}

Unlike competing methods, VEVAR allows the estimation of group-level edges as (possibly) non-linear functions of the covariate values. Figure \ref{fig:EstimatedSim} provides true and estimated subject-level edge strengths and group-level functions, for one of the simulated datasets, and 
Table \ref{Table:Results_by_function} shows TPRs for each of the underlying functions in \eqref{eq:functions} that we used to generate the VAR coefficients, averaged over all replicated datasets. All functions were recovered relatively well for both linear and non-linear functions and the estimated edge strengths were close to the true simulated values. For example, the average MSEs across selected edges generated by function 1 were 0.0034 and 0.0027 for the two groups, respectively, and those for function 3 were 0.0054 and 0.0043, respectively. As expected, it was more difficult to recover complex functions for the group with a smaller sample size, though the performance was generally good. It is also worth noting that the categorical covariate effects were selected correctly at a high rate, as seen in column 5.

\begin{table}
\centering
\begin{tabular}{|l|l|l|l|l|l|l|l|}
\multicolumn{8}{c}{\textcolor{black}{True Positive Rate (TPR)} by Generating Function} \\
\hline
$\mid i - j \mid$ =  & 1 & 2 & 3 & 4 & 5 & 6 & 7\\
\hline
Group 1 & .944 & .909 & .807 & .873 & .999 & .989 & 1.00 \\
Group 2 & 1.00 & 1.00 & .999 & 1.00 & 1.00 & 1.00 & 1.00  \\
\hline
\end{tabular}
\caption{\textcolor{black}{{\bf Simulation study:} Results of the proposed VEVAR method, showing True Positive Rates (TPR) of group-level edge selection for each of the underlying functions in formula \eqref{eq:functions} of Section 2.2, which were used to generate the VAR coefficients. The results, averaged over all replicated datasets, indicate that all functions were recovered relatively well for both linear and non-linear functions.}}
\label{Table:Results_by_function}
\end{table}


\subsubsection{Sensitivity analysis}
\label{sec:sens}
We investigated the sensitivity to the key parameter choices that determine the sparsity of the model, that is $\pi_{\delta}$ and $\pi_{\phi}$. We also considered sensitivity to the choice of $\sigma^2$, the variance of the GP kernel. We used the same simulation setting as outlined above, but with smaller networks ($R = 10$) for computational convenience. Parameters were varied one at a time keeping the others fixed at the values specified in the simulation as described above, and the reported results were averaged across 25 replicated datasets. Results from this sensitivity analysis are shown in Table \ref{table:SimSensitivity}, for both network edge selection and covariate effect selection. The parameter $\pi_{\delta}$ can be interpreted as the prior probability of a group-level VAR coefficient being non-zero. As $\pi_\delta$ increases, so does the number of selected edges. As expected, this parameter had no influence on the selection of the covariate effects. Parameter $\pi_{\phi}$ represents the prior probability of a covariate being selected as relevant for a certain edge strength. As $\pi_\phi$ increases so does the number of selected covariate effects. This in turn causes a slight increase in the number of selected edges. Despite this increase, the performance results remain pretty stable. The parameter $\sigma^2$ represents the variance of the GP kernel. As shown in Table \ref{table:SimSensitivity}, the numbers of selected edges and selected covariate effects increase as $\sigma^2$ increases, but this trend could be the opposite for data with different scales. A general trend that can be noticed in all results is that the inference is more robust to parameter selection for groups with larger sample sizes. 

\begin{table}
\centering
\begin{tabular}{|c|c|c|c|c|c||c|c|c|c|c|}
\multicolumn{11}{c}{{\large{Network Edge Selection}}} \\
\hline 
 & \multicolumn{5}{|c||}{Group 1} & \multicolumn{5}{|c|}{Group 2} \\
 \hline 
 & TPR & FPR & MCC & F1 & Acc & TPR & FPR & MCC & F1 & Acc\\
 \hline
&     \multicolumn{9}{c}{Varying $\pi_\delta$} &       \\
 \hline
  .1 & .733 & .000 & .636 & .845 & .798  & .998 & .010 & .988 & .997 & .996 \\
  .5 & .892 & .460 & .450 & .881 & .814 & 1.00 & .626 & .550 & .907 & .846 \\
  .9 & 1.00 & 1.00 & .000 & .871 & .772   & 1.00     & 1.00 & .000     & .862     & .758  \\
 \hline 
&     \multicolumn{9}{c}{Varying $\pi_\phi$} &       \\
\hline
 .1 & .733 & .000 & .636 & .845 & .798  & .998 & .010 & .988 & .997 & .996 \\
 .5 & .821 & .001 & .739 & .901 & .868  & 1.00     & .018 & .988     & .997     & .995     \\
 .9 & .931 & .000 & .876 & .964 & .948  & .999     & .016 & .987     & .996     & .996 \\
 \hline
 &     \multicolumn{9}{c}{Varying $\sigma^2$} &       \\
\hline
  .1 & .487 & .000 & .429 & .653 & .609  & .928 & .018 & .857 & .959 & .940 \\
  .5 & .651 & .000 & .787 & .786 & .957  & .998 & .008 & .976 & .978 & .994 \\
 1 &.733 & .000 & .636 & .845 & .798  & .998 & .010 & .988 & .997 & .996 \\
 \hline
\multicolumn{11}{c}{{\large{Covariate Effect Selection}}} \\
\hline 
 & \multicolumn{5}{|c||}{Group 1} & \multicolumn{5}{|c|}{Group 2} \\
 \hline 
 & TPR & FPR & MCC & F1 & Acc & TPR & FPR & MCC & F1 & Acc\\
 \hline
 &     \multicolumn{9}{c}{Varying $\pi_\delta$} &       \\
 \hline
  .1 & .535 & .000 & .707 & .692 & .943  & .979 & .002 & .981 & .983 & .996 \\
  .5 & .525 & .000 & .701 & .687 & .941 & .981 & .001 & .984 & .985 & .996 \\
  .9 & .522 & .000 & .697 & .683 & .940   & .981     & .002 & .982     & 0.984 & 0.996  \\
 \hline 
&     \multicolumn{9}{c}{Varying $\pi_\phi$} &       \\
\hline
  .1 & .535 & .000 & .707 & .692 & .943  & .979 & .002 & .981 & .983 & .996 \\
 .5 & .706 & .000 & .824 & .827 & .965  & .999     & .003 & .987     & .988     & .997    \\
 .9 & .882 & .000 & .931 & .936 & .985  & .999     & .004 & .985     & .987    & .997 \\
 \hline
 &     \multicolumn{9}{c}{Varying $\sigma^2$} &       \\
\hline
  .1 & .160 & .000 & .370 & .271 & .895  & .864 & .003 & .906 & .915 & .980 \\
  .5 & .420 & .000 & .602 & .575 & .873  & .923 & .002 & .976 & .978 & .994 \\
 1 & .535 & .000 & .707 & .692 & .943  & .979 & .002 & .981 & .983 & .996\\
 \hline
\end{tabular}
\caption{\textcolor{black}{{\bf Simulation study:} Sensitivity analysis to assess model robustness by varying key parameters and observing their effects on performance metrics. The metrics evaluated are true positive rate (TPR), false positive rate (FPR), Matthew's correlation coefficient (MCC), F1 score, and accuracy (Acc). We focus on the parameters $\pi_{\delta}$ and $\pi_{\phi}$, which determine the sparsity of the model, and $\sigma^2$, the variance of the GP kernel. This analysis helps identify how the model's performance depends on specific input variables.}}
\label{table:SimSensitivity}
\end{table}

\subsection{Results from observational study on traumatic brain injury}
We are interested in estimating group-level connectivity networks and evaluating whether specific edge strengths are affected by covariates of age and sex. For the application of the proposed model, we rescaled the continuous covariate (age) so that the min and max values were -1 and 1. Putting covariates on the same scale is commonly done with GP priors, as it allows the use of the same length scale parameter in the GP kernel function across all covariates.  
We fitted the model to the TBI and HC data using the same parameter settings as outlined in the simulation study, with minor adjustments. Due to the different scale of the fMRI data, we set $\pi_\phi = 0.9$, $\sigma_\mu^2 = 0.01$, and adjusted the variance of the GP kernel function to be 0.5 to better identify the effects of covariates. \textcolor{black}{This prior specification allows us to achieve the expected level of sparsity in brain networks and covariate space and to identify only the most significant connections, therefore retaining interpretability and focusing on the most relevant brain regions and covariates. We are cautious to avoid over-fitting, which could lead to a model too finely tuned to the data. Therefore, the selection of the model is a balance between achieving a sparse, interpretable network and maintaining general applicability. When fitting the model, we use the ELBO to determine the convergence of the variational Bayes algorithm. }

\begin{figure}
\centering
        \includegraphics[width = 0.48\textwidth]{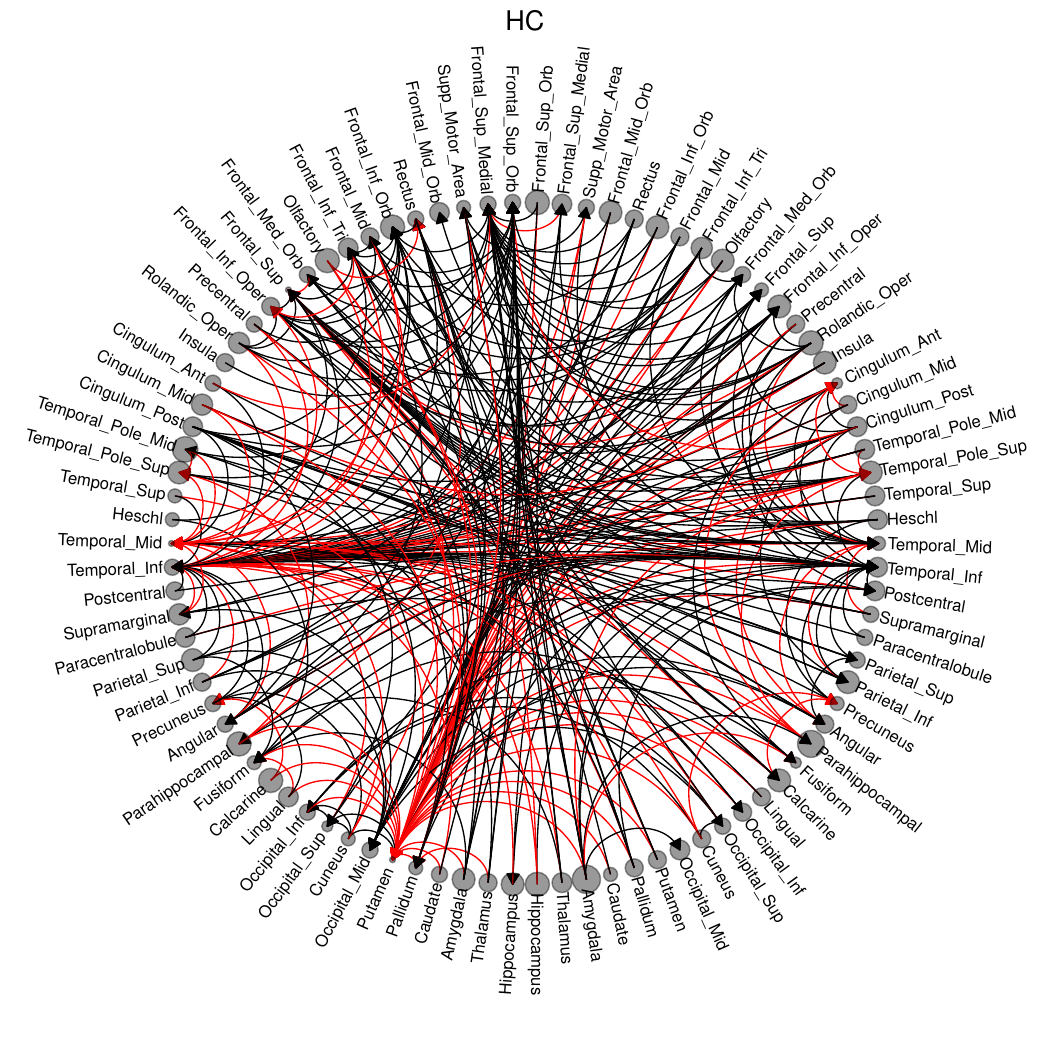}
        \includegraphics[width = 0.48\textwidth]{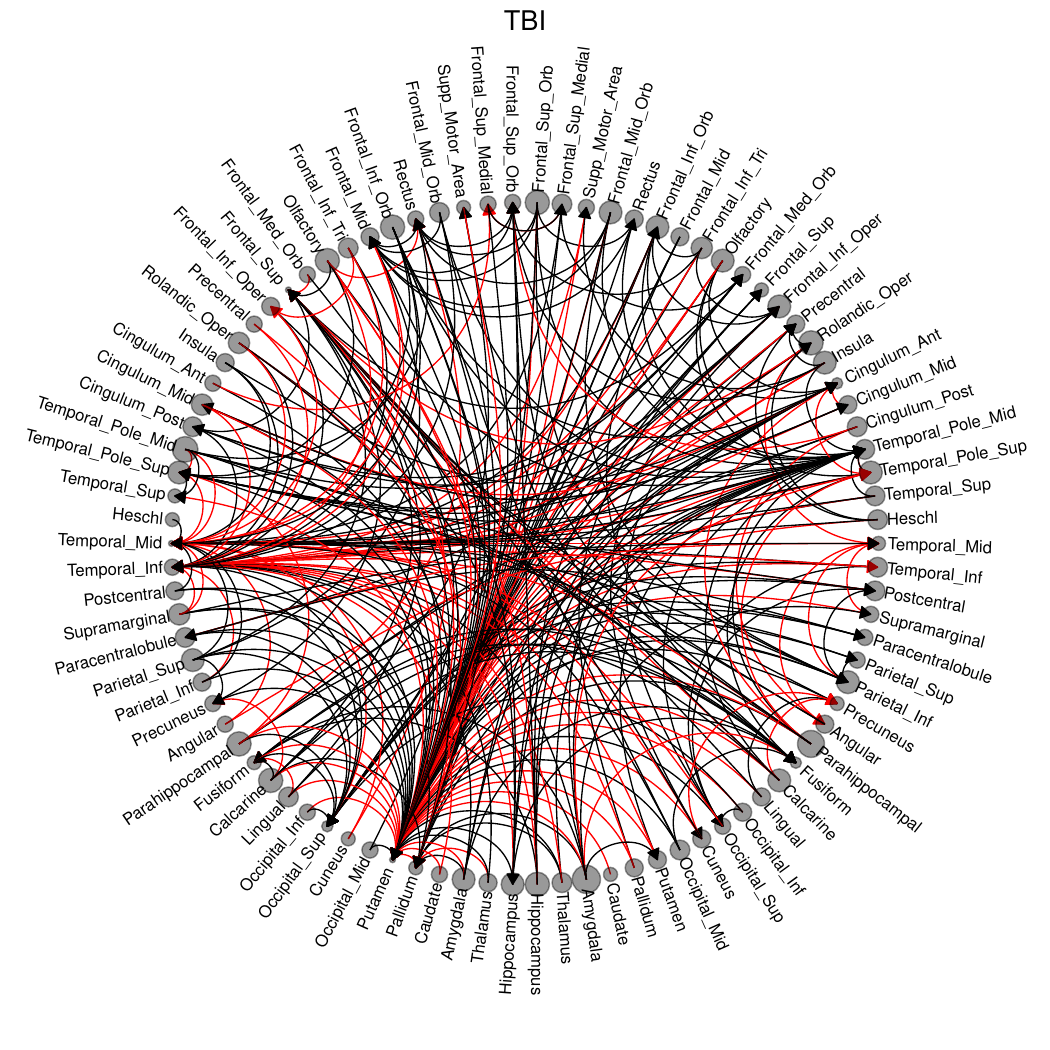}
        \caption{{\bf Application:} Estimated connectograms for HC and TBI groups. Arcs indicate group-level edges (self-connections are not drawn) and node sizes are representative of the number of connected edges, including incoming, outgoing and self-connections. Edges in red are shared between the two groups. \textcolor{black}{The number of selected edges is 419 for the HC group and 405 for the TBI group, with 182 overlapping edges between the two groups. The HC group has more parent nodes than the TBI group in the left frontal lobe regions and right temporal gyrus regions. }}
        \label{fig:connectograms}
\end{figure}

Estimated group-level networks for the two groups, TBI and HC, are shown in Figure \ref{fig:connectograms}, with shared edges in red.  The number of selected edges is 419 for the HC group and 405 for the TBI group, with 182 overlapping edges between the two groups and with the HC group having more parent nodes than the TBI group in the left frontal lobe regions and right temporal gyrus regions. 
\textcolor{black}{We conducted t-tests to compare the estimated subject-level edge strengths between the HC and TBI groups. Our analysis revealed that among the overlapping edges, 154 out of 182 showed significant differences between the two groups. Furthermore, for edges selected in one group but not in the other one, over 95\% of these edges exhibited statistically significant differences in edge strength.}

\begin{figure}
\centering
    \includegraphics[width = 6.5in]{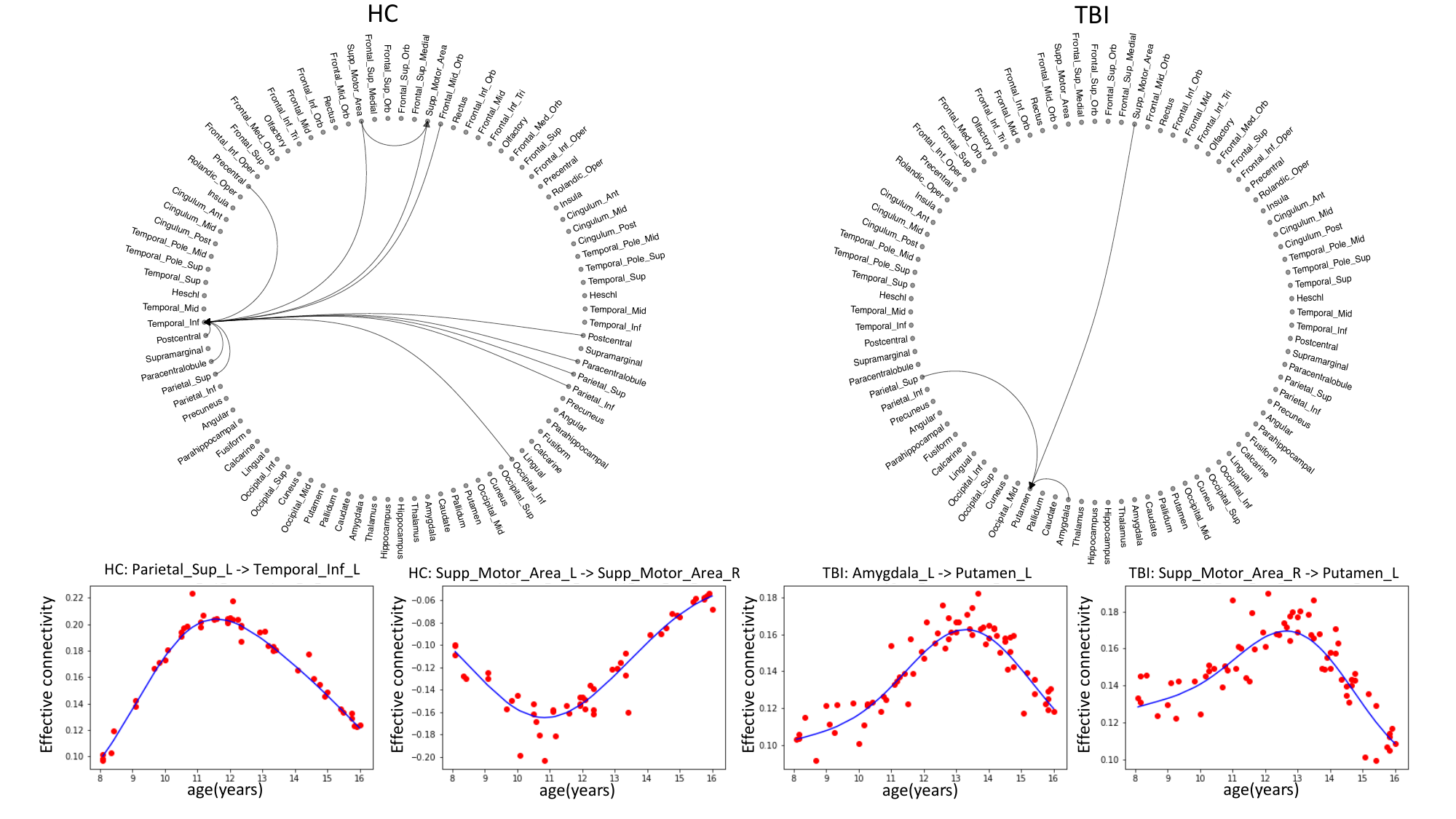}
    \caption{{\bf Application:} (Top) Group-level edges affected by age. An arc between ROIs indicates that the group-level edge is selected and influenced by the covariate `age', \textcolor{black}{with more edges exhibiting age dependence in the HC group.} Self-connections not shown. (Bottom) Subject-level edge strength estimates and the resulting estimated function of the covariate `age' values. Only four of the most interesting estimated effect functions are shown. \textcolor{black}{Notably, the effective connectivity between the left and right supplementary motor areas in the HC group displays a non-linear, inverted U-shaped response to increasing age, initially decreasing then increasing, with a pivot at around 11 years old.}
  }
    \label{fig:CovSelectedAge}
\end{figure}

\begin{figure}
\centering
    \includegraphics[width = 6.5in]{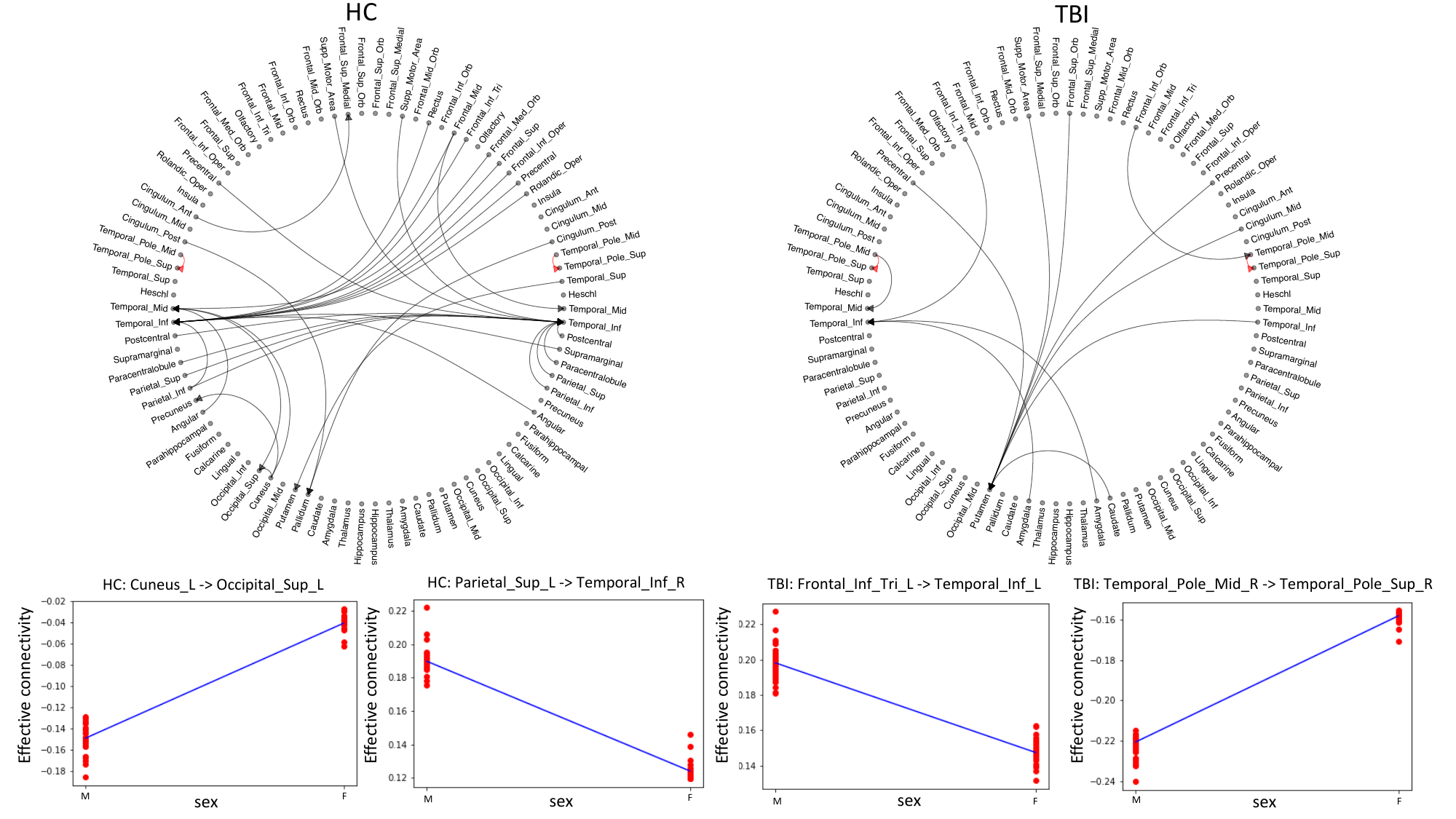}
    \caption{{\bf Application:} (Top) Group-level edges affected by sex. An arc between ROIs indicates that the group-level edge is selected and influenced by the covariate `sex', with edges in red shared between the two groups \textcolor{black}{and more sex-dependent edges in the HC group.} Self-connections not shown. (Bottom) Subject-level edge strength estimates and the resulting estimated function of the covariate `sex' values. Only four of the most interesting estimated effect functions are shown, \textcolor{black}{indicating how edge strengths vary with sex. For example, there is a stronger connectivity in the TBI group between the left triangular part of the inferior frontal gyrus and the left inferior temporal gyrus for males compared to females.}
  }    
    \label{fig:CovSelectedGender}
\end{figure}

Examining the results further, networks displayed in Figures \ref{fig:CovSelectedAge} and \ref{fig:CovSelectedGender} show the selected edge strengths that are affected by age and sex, respectively. From these plots, we see that both age and sex have an influence on the edge strength function for both HC and TBI groups, with more edges in the HC group exhibiting a covariate dependence than in the TBI group. While fewer edges were evident for the TBI group compared to HC, in the TBI group age, and to a lesser extent sex, selectively influenced edges of the left putamen. At a more granular level, our inference reveals how edge strengths change as a function of each covariate. This is illustrated in the bottom parts of Figures \ref{fig:CovSelectedAge} and \ref{fig:CovSelectedGender}, for a few of the most interesting estimated patterns of dependence. For example, in the TBI group, we observe stronger effective connectivity between the left triangular part of the inferior frontal gyrus and the left inferior temporal gyrus for males compared to females. A Table with all selected edges and estimated covariate effect functions can be found in the Supplementary Material, \textcolor{black}{together with plots of the estimated covariate effects for common edges in two groups.} Broadly, we see that the model does well in capturing different functional shapes as well as capturing functions of categorical and continuous data. 

We also compared our findings with three other methods: GC-LASSO, GC-plsmselect, and VEVAR-S1-SS. GC selected 1324 edges for the HC group and 1645 edges for the TBI group, with 101 and 171 overlapping edges with our methods, respectively. VEVAR-S1 selected 398 edges for the HC group and 362 edges for the TBI group, with 215 and 233 overlapping edges with our methods, respectively. 
\textcolor{black}{We note that, as there is no ground truth from real data, the number of selected edges simply illustrates  the relative sparsity of the inferred networks. 
 However, the primary focus of our method lies in its ability to incorporate significant covariates directly into the estimation of edge strengths, thereby enhancing our understanding of individual variability in brain connectivity.} 
Results for covariate effect selections are shown in Table \ref{table:covariateResults-case}. 
\textcolor{black}{Unlike GC, which provides a single estimation for the whole group, our approach allows for a nuanced analysis that can differentiate and track changes across various demographic and developmental stages. Furthermore, GC-LASSO and VEVAR-S1-SS are limited to identifying only linear effects of covariates, while VEVAR can handle non-linear effects.} Although GC-plsmselect is capable of handling non-linear effects, the simulation study results showed that GC-plsmselect had a significantly high FPR, which may explain the larger number of selected edges with covariate effects seen in Table \ref{table:covariateResults-case}.  Finally, it is also worth mentioning that the two-stage methods GC-LASSO, GC-plsmselect and VEVAR-S1-SS may select edges with covariate effects that were not selected in the first stage, which may not be relevant to the study.

\begin{table}
\centering
\begin{tabular}{|l|c|c||c|c|}
      \multicolumn{5}{c}{Covariate Effect Selection}\\
      \hline
      & \multicolumn{2}{|c||}{HC} & \multicolumn{2}{|c|}{TBI} \\
      \hline 
      & Age & Sex & Age & Sex \\
      \hline 
     GC-LASSO &  7 & 23 & 12 & 7\\
     GC-plsmselect  &  55 & 95 & 273 & 105\\
     VEVAR-S1-SS &  2 & 51 & 42 & 17\\
     VEVAR & 26 & 96 & 9 & 75\\
\hline 
\end{tabular}
\caption{{\bf Application:} Results of competing methods and VEVAR for numbers of edges that are influenced by covariates. Note that these results include self-connections within the networks.}
\label{table:covariateResults-case}
\end{table}

\begin{table}
\centering
\begin{tabular}{|l|c|c|c||c|c|c|}
\multicolumn{7}{c}{{\large{Sensitivity Analysis}}} \\
\hline 
 & \multicolumn{3}{|c||}{HC} & \multicolumn{3}{|c|}{TBI} \\
 \hline 
 & Total & Age & sex & Total & Age & sex \\
 \hline
&     \multicolumn{6}{c|}{Varying $\pi_\delta$}      \\

 \hline
  .1 & 419 & 26 & 96 & 405 & 9 & 75\\
  .5 & 458 & 37 & 88 & 434 & 13 & 64\\
  .9 & 527 & 26 & 92 & 507 & 10 & 65\\
 \hline 
 &     \multicolumn{6}{c|}{Varying $\pi_\phi$}      \\
\hline
 .1 & 353 & 24 & 89 & 282 & 13 & 53\\
 .5 & 391 & 29 & 89 & 381 & 13 & 58\\
 .9 & 419 & 26 & 96 & 405 & 9 & 75\\
 \hline
 &     \multicolumn{6}{c|}{Varying $\sigma^2$}  \\
\hline
 .1 & 324 & 13 & 128 & 440 & 14 & 146\\
 .5 & 419 & 26 & 96 & 405 & 9 & 75\\
 1 & 243 & 28 & 47 & 179 & 7 & 36\\
 \hline
\end{tabular}
\caption{{\bf Application:} Sensitivity analysis. Total numbers of selected edges and number of selected edges that are influenced by age and sex.}
\label{table:Sensitivity Case}
\end{table}

\subsubsection{Sensitivity analysis}
Results from the sensitivity analysis for the application conducted on the parameters $\pi_{\delta}$, $\pi_{\phi}$, and $\sigma^2$ are shown in Table \ref{table:Sensitivity Case}. Consistent with the findings from the simulation study, an increase in the value of $\pi_{\delta}$ was observed to correspond to an increase in the number of selected edges. This parameter had no impact on the selection of covariate effects. An increase in the value of $\pi_{\phi}$ led to a rise in the number of selected effects for the binary covariate (sex), while having a limited effect on continuous covariate (age). This increase resulted in an overall increase in the number of selected edges, as we also observed in the simulation study. For the TBI group, our observations revealed a negative relationship between $\sigma^2$ and the number of selected edges associated with covariates. On the other hand, for the HC group with a smaller sample size, no clear trends were observed for the total number of selected edges and the number of selected edges associated with age. This highlights the need for careful selection of the $\sigma^2$ parameter as it can be impacted by both the scales of covariates and the estimated edge strengths. \textcolor{black}{While there are variations in the covariates selected under different parameter settings, the majority of the covariate effects that are identified tend to overlap, and the relationships between effective connectivity and the covariates remain relatively consistent.}

\section{Discussion} 
\label{Discussion}

We have developed VEVAR, a novel extension of a Bayesian vector autoregressive model, and examined its ability to characterize group differences in effective network connectivity while simultaneously evaluating the potentially nonlinear impact of covariates on network edges. Using simulated data sets, VEVAR outperformed other competing approaches in terms of group-level edge selection as well as in covariate effect selection. We then applied VEVAR to assess its utility in characterizing subject and group-level connectivity network edge strengths using resting state fMRI data from a clinical sample of children with TBI and HC. Further, we estimated the effects of age and sex covariates on the group-level edge strengths. The groups differed in the distribution of parent nodes, which were fewer in children with TBI compared to controls. Age effects were largely nonlinear and influenced edges predominantly in healthy children, suggesting alteration in the relation of age with peak edge strength in children with TBI. Group-level edges were also affected by sex; effective connectivity strength was higher in more edges in males than in females. VEVAR has major potential to generate refined analyses of network-level data while clarifying potentially non-linear relations with diverse covariates.

\textcolor{black}{Unlike traditional methods that assume constant edge strengths across the group, VEVAR offers a flexible estimation of network edges that vary with covariate values, which is particularly advantageous in scenarios where edge strengths are potentially related to covariates in a non-linear fashion. For group-level edge selection, VEVAR can model the edge strength as a dynamic function influenced by covariates, rather than as a static feature. In cases where competing methods might overlook an edge due to averaging effects (where the mean edge strength across subjects is zero), VEVAR retains the ability to detect these edges because it accounts for the covariate-driven variability in edge strength. In the simulation study, we compared performances to two-stage approaches that estimate the networks at the first stage and then select the covariates that explain the edge strengths at the second stage. We found that VEVAR does well in terms of all performance measures considered, for both group-level edge selection and covariate effect selection. 
For covariate effect selection, we saw again that VEVAR performs the best across all metrics, demonstrating superior accuracy and sensitivity in identifying the true effects of covariates on network connectivity, likely due to its capability to model both linear and non-linear relationships between covariates and edge strengths.} 
One may raise a concern about comparing VEVAR to other methods for covariate selection, as it was shown that other models did not do well in selecting group-level networks. However, upon examining the subject level edge strength estimates, both GC and VEVAR-S1 gave estimates that were close to the underlying values used to simulate the data, minimizing our concern about this causing the difference in performance.

When applying VEVAR to data from the resting state fMRI study, our findings revealed that in the HC group the left frontal lobe regions and right temporal gyrus regions have more parent nodes than the corresponding nodes in the TBI group. These findings are consistent with those from other analytical approaches and reflect a consensus on the vulnerability of structural and functional connectivity in the frontal and temporal regions following TBI \citep{johnson2011predicting, Vaughn, ware2022structural, watson2019graph, botchway22}. Furthermore, the TBI group showed higher edge strength in the connectivity of the right temporal pole with the right-sided striatal and limbic structures, including the putamen, amygdala and hippocampus. This may reflect the reduced top-down regulation of limbic structures noted after TBI \citep{ewing2019post}.
{\color{black}Regional hyperconnectivity has been reported  in children and adults across the spectrum of TBI severity and chronicity \citep{caeyenberghs2017mapping}.  Hyperconnectivity, especially in the default mode network, has been linked to better task performance in several studies \citep{palacios2017resting, venkatesan2019functional, grossner2019enhanced, stephens2018preliminary, lancaster2019default}. Although hyperconnectivity is often viewed as a compensatory mechanism, the potential long-term metabolic costs have not been established \citep{hillary2017injured}.}

Both age and sex significantly impacted edge strength in HC and TBI groups. In particular, we identified a higher count of edges in the HC group. {\color{black}Edges in the HC group that were influenced by age primarily  connected left inferior temporal to bilateral frontal and temporal-parietal regions.  These anterior and posterior association areas are among the later-developing brain regions. Reduced connectivity in these edges following TBI may reflect vulnerability of these connections among regions that develop rapidly during early to late adolescence.}
We also identified a distinct influence of these demographic factors, particularly age, on the left putamen edges in the TBI group. DTI tractography has previously identified the putamen as a structural hub in children with a history of TBI but not in an age matched HC group \citep{caeyenberghs2012brain}.  Moreover, frontal-striatal (including putamen and caudate) network disruption is evident following pediatric TBI \citep{watson2019graph}. {\color{black} 
Of the few studies examining functional connectivity changes after pediatric TBI, only one evaluated the influence of age and sex on connectivity metrics. In a sample of children with mild TBI, \cite{onicas2024longitudinal} found that by three to six months after injury, global clustering was reduced in  superior parietal and occipital regions in adolescents as well as in females experiencing persistent symptoms relative to orthopedic injury.  
Although our sample contained a broad spectrum of TBI severity,  our findings converge to indicate vulnerability of adolescents and females to disruption of specific connectivity metrics.}
Current findings elucidate the functional implication of striatal injury and characterize how patient characteristics, in this case, age or sex, might strengthen or attenuate the impact of injury on striatal brain networks. 

The effective connectivity between the left supplementary motor area and right supplementary motor area for the HC group have shown an interesting non-linear, inverted U shape as age increases, first decreasing and then increasing. The trends of the estimated effective connectivities with respect to age change at around 11 years old for the HC group while the trends change at around 13 years old for the TBI group. Importantly, our approach does not constrain associations between covariates and edge strength to a non-linear function but instead allows for discovery of the pattern of influence of the covariate to be detected. 
Developmental profiles showing inverted U-shapes are consistently reported in brain morphometry via regional volume or thickness of cortical gray matter and are broadly interpreted to reflect initial overly abundant synaptic production and the subsequent synaptic pruning \citep{giedd1999brain, giedd2006puberty, gogtay2004dynamic}.  Although functional profiles corresponding to an inverted U during childhood and adolescence are established in the literature, it is to a lesser degree relative to the morphometric literature. Notably, the pattern of inverted U in brain functioning, even in development, is most often associated with an increase in expertise or skill (often in language/reading domains), such that learning initially triggers an increase (the rise in the inverted U) with the falling end of the U associated with experience, and ultimately expertise, see \cite{perkins2019neuroimaging}. 

Despite its importance as a biological variable, few imaging studies have examined how sex may affect connectivity metrics. In our results, many group-level edges were influenced by sex. Except for bilateral temporal pole regions, very few edges were shared in females and males. Relations also differed by group. In the TBI group, females had higher effective connectivity than males in both temporal poles. Males had a unique pattern wherein edges originating in multiple bilateral structures converged in the left putamen. Effective connectivity in males in the HC group was higher in multiple bilateral regions impacting bilateral mid- and inferior temporal regions while females had higher connectivity in edge strengths affecting left temporal regions. \textcolor{black}{We note that there could be other confounding factors, such as head size, influencing the observed sex differences in brain connectivity. Given that we do not have head size data, we cannot conclusively link the correlations directly to sex. Therefore, our results should be viewed as indicative of a correlation between sex and specific brain connectivity patterns, rather than conclusive evidence of sex-based differences.} Additional studies should examine the relation of connectivity metrics and covariate effects with cognitive and behavioral outcomes to more fully elucidate specific brain-behavior relations.

\textcolor{black}{There are several aspects of the proposed model that allow the practitioner added flexibility. For example, as previously discussed, the GP kernel can differ across datasets, edges, or covariates. 
In both the simulation and case study application, the covariates used were subject-specific measurements, constant across edges and nodes. The model, however, could easily be adjusted to accommodate covariates specific to a subset of edges or groups. Alternatively, covariates whose values vary at each node or edge could be considered. Finally, while the current model formulation relies on the group subject memberships to be known, unsupervised settings could also be considered via nonparametric priors such as the Dirichlet process. This would lead to broader applications and potentially further insights into the data.}

\section{Conclusion}
\label{Conclusion}
In this paper, we have proposed VEVAR, an analytical approach for estimating brain connectivity networks that accounts for subject heterogeneity. We have employed a hierarchical Bayesian varying-effects vector autoregression model to identify connectivity networks for different groups of subjects, allowing for dependence of the connection strength on covariate values. We have designed a sparse prior to identify key connections within each group and subsets of edges with strengths affected by the covariates. Our novel nonparametric spike-and-slab prior includes a ``slab" portion as a function mapping possible covariate values to coefficient values. Additionally, we have assumed a weighted mixture of Gaussian process priors on this function, to allow modeling of (possibly) non-linear effects and selection of relevant covariates. We have estimated the proposed model using variational inference, which has allowed for application to large-scale data. The model has been shown to perform better than competing two-stage approaches on simulated data, in both network discovery and covariate selection.  We have applied our method to resting-state fMRI data on children with a history of TBI and healthy controls to estimate group-level connectivity networks and evaluate whether specific edge strengths are affected by age and sex. The identified differences in effective connectivity network patterns between TBI and HC groups, along with the influence of age and sex, have provided valuable insights into the complexity of brain functioning after TBI. Furthermore, TBI as a case study application, where effects of age are evaluated within effective connectivity networks, highlights the promise of the proposed approach for evaluating developmental and aging effects in a breadth of disorders impacting the brain. For future studies, extensions to longitudinal data may provide valuable insights into connectivity changes over an extended period after TBI.

\textcolor{black}{Broadly, VEVAR contributes to the current understanding of brain mechanisms by identifying functionally connected brain regions, enabling researchers to characterize information flow within and between brain networks. Moreover, VEVAR determines the directionality of functional connections between brain regions, offering insights into how specific cognitive functions like memory, attention, language, and decision-making are temporally implemented within overlapping brain networks. In the context of development, VEVAR advances understanding by tracking the development of neural networks from infancy through adolescence and into adulthood. By observing changes in connectivity patterns over time, researchers can identify sensitive developmental periods and map the integration of different brain regions. VEVAR in brain connectivity research has been implemented to provide crucial insights into altered neural circuits associated with injuries such as traumatic brain injury \citep[TBI; ][]{kook2021bvar, vaughn2022effective}. This method is poised for similar use in neurodevelopmental disorders like autism spectrum disorder \citep[ASD; ][]{hanson2013atypical, rolls2020effective}, dyslexia \citep{di2023disentangling}, or attention deficit hyperactivity disorder \citep[ADHD; ][]{kumar2021neural}, where effective connectivity findings suggest that differences in the connectivity patterns in individuals with developmental disorders compared to typically developing individuals. Moreover, VEVAR facilitates the comparison of connectivity profiles between typically developing individuals and those affected by these disorders while controlling for nuisance variables. This enables analyses that may identify aberrant developmental trajectories and potential biomarkers for early diagnosis and intervention.}


\section*{Acknowledgments}
Y.R., M.V. and C.B.P were supported in part by grants DMS-2113602 and DMS-2113557 from the National Science Foundation.   D.M.D. and L.E.C. were funded in part by National Institutes of Health R01 NS046308. The content  is solely the responsibility of the authors and does not necessarily represent the official views of the granting institutes.

\section*{Competing interests}
The authors declare no competing interests.

\section*{Data availability}
The processed imaging data analyzed in this paper can be downloaded from the MATLAB BVAR package at 
https://github.com/marinavannucci/BVAR$\_$connect. The subject-level covariate information that supports the findings of this study is available from the corresponding author upon reasonable request.

\section*{Code availability}
Python code implementing the VEVAR model described in this paper can be downloaded from GitHub at \textcolor{black}{https://github.com/YangfanR/VEVAR-fmri}.

\bibliographystyle{apalike}
\bibliography{ref}

\end{document}